\documentclass{pasj00}

\Received{$\langle$reception date$\rangle$}
\Accepted{$\langle$acception date$\rangle$}
\Published{$\langle$publication date$\rangle$}

\SetRunningHead{Astronomical Society of Japan}{Usage of \texttt{pasj00.cls}}

\usepackage{times}

\begin{document}

\title{V369 Gem: A New Ellipsoidal Variable System \\
With High Level Chromospheric Activity}

\author{Dal, H. A., Sipahi, E., \"{O}zdarcan, O.}
\affil{Department of Astronomy and Space Sciences, University of Ege, \\
Bornova, 35100 ~\.{I}zmir, Turkey}

\email{ali.dal@ege.edu.tr}


\maketitle

\begin{abstract}

In this study, we present BVR photometry of V369 Gem in 2006 - 2008. We combined our V data with the data taken from the literature. We analysed all the data together and calculated the new light elements. General features of the light curves were discussed. The light curves were modelled with the Fourier method and the Fourier coefficients calculated. The $\cos (2 \theta)$ term was found to be dominant. The pre-whitened V light curves demonstrated that stellar spot activity also exist. There are generally two minima in the pre-whitened light curves. Thus, there must be two spotted areas, which are separated about 180$^{\circ}$ from each other. As conclusions, we have remarked for the first time that V369 Gem is an ellipsoidal variable with the high level chromospheric activity. \\

\end{abstract}

\section{Introduction}

Many stars from the spectral types G, K and M exhibit solar like chromospheric activity. The stellar spots on the photosphere of a star cause quasi-periodic sinusoidal-like variations due to the star's axial rotation. In this case, the long-term photometric monitoring of active stars provides a powerful tool to derive relevant parameters of stellar surface activity \citep{Rod00, Mes03}. In this study, we obtained the light curves of V369 Gem (HD 52452) in 2006-2008 years. Although V369 Gem is known as one of the short-period non-eclipsing chromospherically active binary system discovered so far \citep{Mes01}, we investigate the main reasons of the light variations seen in its light curves due to some stable properties seen in the light curves.

In literature, for the first time, \citet{Pou93} detected the star as a EUV bright source in the all sky survey of ROSAT. They listed the star as RE J70222+255054 in the X-Ray source catalogue. Later than, \citet{Mas95} demonstrated that V369 Gem is the optical counterpart of the EUV bright source RE J70222+255054. \citet{Per97} catalogued the star as a suspected variable star with a parallax of 17.0$\pm$6.4 mas in the Tycho Catalogue (TYC1899 688 1). \citet{Mes01} reported that V369 Gem is a triple system. They indicated that the triple system is composed of a G4V star and two late type G stars, which are seen as SB1. The observations of the authors indicated that both components of the binary system exhibit photospheric spot activity. The authors stated in their study that there was no proof of an eclipsing in the light variation. They derived an upper value as 50$^{\circ}$ for inclination of the SB1 components. It was mentioned that the system is not young in terms of the abundance analyses of Li $\lambda$6707$\AA$ in their spectral studies with high resolution. The $v \sin i$ value is given as 14$\pm$2 kms$^{-1}$ for one of the components, while $v \sin i$ value of the other component is higher than 60 kms$^{-1}$. For the first time, photometric period of the system was found to be 0.42304 days by \citet{Mes01}. Another study belongs to \citet{Bar04}, who presented BVRI-band photometry of the system. The authors gave the photometric periods, which they found out by analyses of their data together with the data given by \citet{Mes01}. The authors mentioned that the light variation of the system was due to a stellar spot activity on the components of the system. They reported that the amplitudes of the light curves were changing in 2000 and 2001 observing seasons. Using the data taken from the Tycho Catalogue, \citet{Bar04} indicated that the distance of the system was 59 pc.

In this study, we suggest that there are two main effects to cause the light variation. Examining all the light curves of the system, one of them should be the ellipsoidal effects, which cause a stable light variation with two minima. The second one is the chromospheric activity, which cause the short-term rapid variation on this stable shape. In the paper, we analysed all the V-band data of the system acquired between the years 1995 and 2009. The observational bases of the study are presented in Section 2, then photometrically obtained light variations and analyses are presented in Section 3, Section 4 and Section 5. Finally, the main properties obtained in this study and the results are discussed in Section 6.

\section{Observations and Data}

The observations were acquired with a High-Speed Three Channel Photometer attached to the 48 cm Cassegrain telescope at Ege University Observatory. The observations were continued in BVR-bands. The identities of programme star and its comparisons are given in Table 1. In Table 1, the star names are given in the first column, while J2000 coordinates are listed in second column. The V magnitudes and B-V colours obtained in this study are listed in the following columns, respectively.

V369 Gem, the comparison and check stars are close on the sky plane. However, differential atmospheric extinction corrections were applied for the observations obtained in this study. In each observing night, the observations of the system were acquired without interruption, covering large ranges for the airmass. In general, the observations were started before the celestial meridian with the airmass value of 3.15, and were ended after the celestial meridian with the airmass value of 2.25. The coefficient ($k$) of differential atmospheric extinction correction was found to be between 0.224 and 0.390 $mag~airmass^{-1}$ for B-band observations. The mean value of $k$ was computed as 0.321 $mag~airmass^{-1}$ for B-band observations. It was found to be between 0.085 and 0.299 $mag~airmass^{-1}$ for V-band, while its mean average is 0.221 $mag~airmass^{-1}$. Finally, it was found to be between 0.038 and 0.274 $mag~airmass^{-1}$ for R-band observations, and its mean value was computed as 0.149 $mag~airmass^{-1}$. The time series analyses also do not reveal any regular variation for these differential magnitudes. There is no variation of differential magnitudes in the sense comparison minus check stars. The comparison stars were observed with the standard stars also in their vicinity and the reduced differential magnitudes, in the sense variable minus comparison, were transformed to the standard system using procedures outlined in \citet{Har62}. The standard stars we used are listed in the catalogue of \citet{Oja96}. The derived transformation coefficients are given in Equations (1, 2 and 3). Heliocentric corrections were also applied to the times of observations. The standard deviations of observations acquired in BVR-bands are 0$^{m}$.010, 0$^{m}$.007 and 0$^{m}$.006, respectively.

\begin{center}
\begin{equation}
\label{eq:one}
(V-v_{0})~=~-0.0113~\times~(b-v)_{0}~+~20.877
\end{equation}
\end{center}

\begin{center}
\begin{equation}
\label{eq:two}
(B-V)~=~-0.9924~+~1.1592~\times~(b-v)_{0}
\end{equation}
\end{center}

\begin{center}
\begin{equation}
\label{eq:three}
(V-R)~=~-0.0698~+~1.079~\times~(v-r)_{0}
\end{equation}
\end{center}
where the symbols $bvr$ represent the observed values, while the symbols $BVR$ represent the standard values.

In this study, we used the same comparison star with both \citet{Mes01} and \citet{Bar04}. \citet{Mes01} given the observation data in the standard system, while \citet{Bar04} given the values of the comparison star in their study. The comparison star brightness and colour given by \citet{Bar04} are close to the values obtained in this study. The V data obtained by \citet{Mes01} and \citet{Bar04} were combined to our data. In addition, we also used the data obtained from the ASAS Database \citep{Poj97}. The system was observed between December, 2002 and April, 2009 in the ASAS project. In Figure 1, all the V-band data were plotted versus years.

All the V-band data were analysed with both Discrete Fourier Transform (DFT) \citep{Sca82} and the Phase Dispersion Minimization (PDM), which is a statistical method \citep{Ste78}. In the analyses, the period was found to be 0$^{d}$.423061. Using the method given by \citet{Kwe56} and the data obtained in January 10, 2007 night, an epoch was determined. The new light elements are given in Equation (4).

\begin{center}
\begin{equation}
\label{eq:four}
JD~(Hel.)~=~24~54111.5034(4)~+~0^{d}.423061(1)~\times~E
\end{equation}
\end{center}

All the combined data were phased with the light elements given in Equation (4). Considering that V369 Gem is an active star, the consecutive observations were compared among themselves. The date, where the shape and the amplitude are changed, was taken as the beginning of the new subset. Thus, all the V data were separated into 14 subsets. Some parameters of the subsets are listed in Table 2. In Figure 2, the BVR light curves of the system obtained in this study are given for only two subsets, which are covered all the phase intervals. In Set L and Set M, there are also some data obtained in this study, but these data are not enough to cover all the phases. This is why we did not plotted them in Figure 2. As it is seen from the figure, the shapes and amplitudes of the light curves different especially in minima phases.

\section{Light Curve Analysis}

According to \citet{Mes01} and \citet{Bar04}, the observed light variation is caused due to the cool spots in the case of V369 Gem. However, there are a few important characteristic features in the light curves. First of all, there are always two minima in the light curves and one of them is always deeper, and there is 0$^{P}$.50 between them. In addition, the phases of the minima are generally constant, though there could be some small distortions in the general shape of the light curves time to time.

Considering these properties, we tested whether there are any eclipses in the system by using the PHOEBE V.0.31a software \citep{Prs05}, whose method depends on the version 2003 of the Wilson-Devinney Code \citep{Wil71, Wil90}. We tried to analyse BVR-band curves with different modes, such as the "detached system", "overcontact binary not in thermal contact", "semi-detached system with the primary component filling its Roche-Lobe", "semi-detached system with the secondary component filling its Roche-Lobe" and "double contact binary" modes. Considering the minimum sum of weighted squared residuals ($\Sigma res^{2}$) of 0.13, the initial analyses demonstrated that a statistically acceptable result can be obtained if the analysis is carried out in the "overcontact binary not in thermal contact" mode. The BVR-light curves can not be modelled by using all the other modes. In the case of "detached system" mode, the analysis is generally interrupted itself due to uncomfortable configuration, while the program is working. In the case of "double contact binary" mode, the obtained $\Sigma res^{2}$ values are much larger than those obtained by using the "overcontact binary not in thermal contact" mode. Because of these, an available solution was obtained in just the "overcontact binary not in thermal contact" mode.

Using spectral observations, \citet{Mes01} estimated that the system is G spectral type, and so the effective temperature ($T_{eff}$) of the primary component should be 5637 $K$ according to \citet{Dri00}. Thus, the temperature of the primary component was fixed at 5637 $K$, and the temperature of the secondary was taken as a free parameter in the analyses. Considering the spectral type, the albedos ($A_{1}$ and $A_{2}$) and the gravity-darkening coefficients ($g_{1}$ and $g_{2}$) of the components were adopted for the stars with the convective envelopes \citep{Luc67, Ruc69}. The non-linear limb-darkening coefficients ($x_{1}$ and $x_{2}$) of the components were taken from \citet{Van93}. In the analyses, the fractional luminosity ($L_{1}$) of the primary component and the inclination ($i$) of the system were taken as the adjustable free parameters. There is no available spectroscopic mass ratio for the system. Because of this, we tried to find the best photometric mass ratio of the components using the light curve analysis. The minimum sum of weighted squared residuals ($\Sigma res^{2}$) obtained in the analyses indicated that the mass ratio value of the system is $q= 0.76\pm0.03$. The synthetic light curves obtained in the "overcontact binary not in thermal contact" are shown in Figure 3, while the parameters derived from this analysis are listed in Table 3. From the parameters found from the light curve analysis, the 3D model of V369 Gem's Roche geometry is shown in Figure 4 with the geometric configurations of the system at four special phases 0.00, 0.25, 0.50 and 0.75.

According to the minimum sum of weighted squared residuals ($\Sigma res^{2}$) of 0.13, the only solution obtained in the "overcontact binary not in thermal contact" is statistically acceptable one. However, it should be tested whether it is also acceptable in the astrophysical sense. For this aim, although there is not any available radial velocity curve, we tried to estimate the absolute parameters of the components in order to compare them with the theoretical models. According to \citet{Dri00}, the mass of the primary component must be 0.946 $M_{\odot}$ corresponding to its surface temperature. Considering possible mass ratio of the system, the mass of the secondary component was found to be 0.721 $M_{\odot}$. Using Kepler's third law, we calculated the semi-major axis as 2.81 $R_{\odot}$. Considering this estimated semi-major axis, the radius of the primary component was computed as 1.06 $R_{\odot}$, while it was computed as 0.92 $R_{\odot}$ for the secondary component. Using the estimated radii and the obtained temperatures of the components, the luminosity of the primary component was estimated to be 1.022 $L_{\odot}$, and it was found as 0.394 $L_{\odot}$ for the secondary component. If the obtained absolute parameters are compared with theoretical models such as one developed for the stars with $Z=0.02$ by \citet{Gir00}, it is seen that the absolute parameters are generally acceptable in the astrophysical sense.

\section{The Ellipticity Effect}

The light curve analyses demonstrated that the system do not exhibit any eclipses. However, as seen from the Roche geometry shown in Figure 4, the shape of the components should be distorted due to tidal effects. Thus, the ellipticity effect due to this geometric configuration should cause the observed light variation. As it is well known, the ellipsoidal variable stars are non-eclipsing binary stars and especially close binaries \citep{Mor85, Bee85}. The main variation seen in their light curves is due to non-spherical shapes of the components. This like configuration of a system creates a quasi-sinusoidal light curves. The depths of the minima in the light curve depend on the system geometry as well as its inclination.

The one of the methods to test whether a star is an ellipsoidal variable is the Fourier analysis \citep{Mor85, Bee85}. It has been also demonstrated by \citet{Mor85} and \citet{Mor93}, in the case of any available radial velocity (a double lined or single lined), many parameters of the components can be computed under some assumptions. Unfortunately, there is no available radial velocity for V369 Gem; because of this case lots of parameters can not be computed. As it is listed in Table 3, the inclination angle ($i$) of the system was determined as $i=44^{\circ}.25$ from the light curve analysis.

In this case, the stable-main variation could be caused by the ellipsoidal effects. In order to test this, the light curves of the ellipsoidal variables can be modelled by the Fourier analysis given by Equation (5). If the main variation is caused due to the ellipsoidal effect, one will expect the $\cos (2 \theta)$ term must be dominant among all other terms in the results of the Fourier analysis.

\begin{center}
\begin{equation}
\label{eq:five}
L(\theta)= A_{0} ~ + ~ \sum_{\mbox{\scriptsize\ i=1}}^N ~ A_{i} ~ cos(i \theta) ~ + ~ \sum_{\mbox{\scriptsize\ i=1}}^N ~ B_{i} ~ sin(i \theta)
\end{equation}
\end{center}

The light curve obtained in January 10, 2007 was chosen to analyse. The observations in this night cover homogeneously all the phases, and the minima in the light curve are symmetric. The BVR light curves obtained in January 10, 2007 were analysed with the Fourier method. The derived Fourier fits are shown in Figure 5, and the Fourier Coefficients are listed in Table 4. The $A$ coefficients listed in the table are the coefficients of the $\cos(i \theta)$ terms, while $B$ parameters are the coefficients of the $\sin(i \theta)$ terms given in Equation (5). In fact, the most dominant one is $\cos (2 \theta)$ term for each of the BVR-bands. Thus, it is obvious that the main effect seen in the light variations is the ellipticity effect. In this case, the found period given in Equation (4) must be the orbital period of the system. As it is well known from eclipsing binaries, the found photometric period must be the orbital period of the system, if the main reason of the light variation is the geometric effects of the binary. There are many systems, which are similar to V369 Gem, such 75 Pegasi and 42 Persei are studied with this method by \citet{Mar90, Mar91}.

\section{Stellar Spot Activity}

In the Fourier Coefficients, the second dominant term is $\cos (\theta)$. \citet{Hal90} discussed that the $\cos (\theta)$ term can be an indicator for the spotted area on the surface of a star. In this case, the ellipticity effect is not unique effect in the light variation of V369 Gem. Considering the variation seen in Figure 2 and spectral type of the system, the second effect should be stellar spot activity. We extracted the theoretical light curve, which contains just the variation caused by the ellipticity effect, from all the V-band light curves of 14 subsets. We examined all the pre-whitened light curves together, and some parameters were calculated and given in Table 5. All the observed and pre-whitened light curves are shown in Figure 6. The curves seen in the first two columns of panels on the left side are observed light curves, while the curves in the last two columns of panels on the right side are the pre-whitened light curves. In the first two columns of the panels, the dashed line in each panel represents the theoretical light curve, which contains just the variation caused by the ellipticity effect. However, in the last two columns of the panels, the dashed line in each panel represents the theoretical light curve, which contains just the variation caused by the stellar spots.

The variation of the mean brightness in the pre-whitened light curves are shown in Figure 7 with the variation of the amplitudes in the light curves. The pre-whitened V light curves were analysed by time series analyse methods. The found periods are listed in Table 6. Excepted a few subsets, which have no good data distribution, the periods found from each subset are close to the period given in Equation (4). This case reveals that the components rotate synchronously with the orbital period. The data used in analyses are not enough to discuss about any activity cycle.

In Figure 8, we plotted minima phases ($\theta_{min}$) derived from the pre-whitened light curves versus the years. The figure reveals two main properties of the spot distribution on the surface. As seen from the figure, the minima of pre-whitened V light curves are separated 0$^{P}$.50 from each other. This indicates that the spotted areas on the surface are generally inclined to be located in two mean active longitudes. Moreover, the deeper minimum migrates toward the earlier phases through years.

\section{Results and Discussion}

The goal in this study is determining the nature of this interesting system. In this respect, we analysed almost all the available data of V369 Gem in the literature. The results show some clues about its nature.

As seen in Figure 2, the light curves of the system vary from a season to the next one. There are also some variations in the mean brightness and the shapes of the light curves. The same cases were seen by \citet{Mes01} and \citet{Bar04}, and they explained this situation with the chromospheric activity.

We analysed all the V data together with the time series analyses, and the photometric period of the system was found to be 0$^{d}$.423061. It is so close to the periods found by \citet{Mes01} and \citet{Bar04}.

According to the long-term photometry, another properties have come out in this case. Although there are some small variations, the main shapes of the light curves are usually the same. There are always two minima, and they are almost constant according to each other. The analyses demonstrated that one of the effects on the light curves is the ellipsoidal effect. The $\cos (2 \theta)$ was found to be -0.0400$\pm$0.0004 for B, -0.0383$\pm$0.0004 for V and -0.0358$\pm$0.0003 for R-band. These coefficients are larger than all other coefficients in each band. According to \citet{Mor85} and \citet{Hal90}, in this case, there is an ellipsoidal effect on the light variations. Apart from the $\cos (2 \theta)$ coefficients, the second dominant coefficient is $\cos (\theta)$. According to \citet{Hal90}, if there was only one spotted area on the surface of a star, it would be expected that the $\cos (\theta)$ term must be dominant. However, \citet{Hal90} noticed that the $\cos (2 \theta)$ term is not be dominated due to only the ellipticity effect, but it can be also dominated because of two spotted areas separated 180$^{\circ}$ from each other on the surface of a component. However, there is a way to define the real reason of the $\cos (2 \theta)$ term dominated in the Fourier analysis. This is the evolution and movements of the spotted areas on the surface of the star. If the reason of the $\cos (2 \theta)$ term is stellar spots separated 180$^{\circ}$ from each other, the results of the Fourier analyses will change with time. Because, the main shapes of the light curves will change. In this study, we always found the $\cos (2 \theta)$ term to be dominated for each set. This demonstrated that the main effect on the light variation is the ellipsoidal effect. However, one can suspect that V369 Gem may be an eclipsing binary. In this point, the suspect can be tested by the method described by \citet{Mor85} and \citet{Mor93}. In the case of V369 Gem, although there is not any available radial velocity, we determined the inclination angle ($i$) of the system from the light curve analysis with the PHOEBE V.0.31a software, and it was found to be 44$^{\circ}$.25. In this respect, it is not expected that the system exhibits any eclipses.

In order to the find the second effect on the light variation, the ellipsoidal effect was extracted from all the observed V light curves. A sinusoidal-like variation is seen in all the pre-whitened light curves, and this variation seems to be a combination of two sinusoidal waves. The consecutive light curves demonstrated that the shapes of the pre-whitened light curves are dramatically changing. For example, if the pre-whitened light curves of Set C, Set D and Set E are compared, it can be seen that the level of the deeper minimum is increasing, while the level of the first maximum seen in $\sim$0$^{P}$.40 is not changing. In contrast to the first maximum, the level of the second maximum seen in $\sim$0$^{P}$.90 is increasing and, it gets a higher level than the first one in a few subsets. The same behaviours are seen among the others. Although the mean brightness of the pre-whitened light curves is increasing through years, the amplitudes are decreasing from the year 1994 to 2007. The amplitude of the pre-whitened light curve is 0$^{m}$.160 in the first light curve, while it is slowly decreasing toward the year 2007, and it is 0$^{m}$.012 in 2007. However, the largest amplitude is seen in the year 2009.15, it is as large as 0$^{m}$.206. There is one minimum in some light curves, which are generally obtained between the year 2006 and 2007. The mean brightness of the subsets are changing. For instance, the amplitudes of the first six light curves are changing from one to the next, though two minima always exist in the light curves. However, the shallow minimum is beginning to disappear from Set G, and it is disappeared in Set H. The light curve of Set H has one minimum and completely asymmetric shape. The second minimum is started to appear in the light curve of Set I, and the light curve of Set J has two minima. The light curve of Set K has again one minimum and an asymmetric shape. Although the data in both Set L and Set M do not cover all the phases of the light curves, an asymmetry can be seen in their shapes. Finally, the light curve of Set N has two minima. Besides the second minimum sometimes disappearing, another dramatic variation is seen in the light curves. As seen from Figure 8, there is a migration of the deeper minimum. The deeper minimum migrates toward the earlier phases through years, especially the last few years. The sinusoidal-like variations in each light curve should be caused by dark stellar spots occurring on the surface of the system. In brief, the behaviours like these are generally seen in chromospherically active stars \citep{Fek02}.

The minima of the sinusoidal-like variation in the pre-whitened light curves are generally separated about 0$^{P}$.50 from each other. This reveals that there are two spotted areas separated $\sim$180$^{\circ}$ from each others. A sudden occurring of the deeper minimum between 0$^{P}$.70-0$^{P}$.80 in Set H should be caused by changing of the active area efficiencies on the surface of the star. Changing of the active area efficiencies exhibits itself in Figure 6. As seen from the figure, the phases of the maxima can occur in different phases, which are separated 0$^{P}$.50 from each other. The same behaviours are generally seen in the light curves of many close eclipsing binary systems, such as W UMa type variables, but not limited to this type \citep{Dav84, Kal09}. In the W UMa systems, the chromospheric activity occurring on the surface of the components is one of the reasons causing the phenomenon called the O'Connell Effect.

After the extraction of the ellipsoidal effect from the light curves, the periods found from each subset are a bit different from each other. This case is common for active stars. These photometric periods depend on the location(s) of the spotted area(s) on the surface of the star. In fact, if the subsets, which have scattered data, are ignored, the periods found from each subset seem to decrease from the year 1994 to 2009. On the other hand, the similarity between the orbital period and the periods found from each subset demonstrates that the components rotate nearly synchronously with the orbital period. In fact, the rotational and revolution periods are expected to be same for an ellipsoidal variable \citep{Mor85}. In the case of V369 Gem, although the orbital period was found to be close to the period of rotational modulation, it is seen from Figure 8 that there should be a bit difference between them. The minima of the pre-whitened light curves migrate toward to the previous phases.

V369 Gem has been taken a special place among the other ellipsoidal variables. This is because the system has chromospheric activity. In fact, there are many systems similar to V369 Gem in the literature, such as V350 Lac, V1197 Ori, V1764 Cyg, HD 74425 and ect \citep{Cre95, Lin87, Hen99, Hal90}. As it is seen from the literature, although the ellipsoidal variables are generally from the earlier spectral types, but there are also several ellipsoidal systems from the later spectral types. Considering these ellipsoidal variables from the later spectral types, V369 Gem is a bit different among them due to its period. As seen from \citet{Hal90}, they are generally long period systems, while V369 Gem is a very short-period one. The short period of the system indicates that V369 Gem is not a classical ellipsoidal variable, whose shapes are just caused by the tidal effect. The ellipsoidal shapes of V369 Gem's components should be due to not only tidal effects but also the evolutionary status of the system. Although V369 Gem is mentioned as a main sequence star, as seen from Figure 4, the components are near the filling Roche-Lobes. Considering the obtained absolute parameters of the components with the theoretical models derived for the for stars with $Z=0.02$ by \citet{Gir00}, both of them seem to be closer to TAMS rather than ZAMS. \citet{Mes01} demonstrated that very low abundance of Li $\lambda$6707$\AA$ is obtained from this system, because of this, the components can not be the young stars, whereas the system has a very high level chromospheric activity. This is in agreement with our results obtained from the light curve analysis. In fact, \citet{Roc02} put forwarded some models about the formation of W UMa type systems, indicating some systems, which are chromospherically active without high Li abundance, while they are kinematically old. According to \citet{Roc02}, the components in these systems are also generally rapidly rotating. \citet{Mes01} revealed that the components of V369 Gem are also rapidly rotating according to a star from the G spectral types. Consequently, it is possible that the components of the system should be nearly filling their Roche-Lobes. This also explains why the components have taken an ellipsoidal shapes.

In the future, V369 Gem should be taken into the programs of the spectroscopic observations. A radial velocity curve of the system will be obtained, and this supports to understand the unclear points about the system.

\section{Acknowledgments} The authors acknowledge generous allotments of observing time at the Ege University Observatory. We also thank the referee for useful comments that have contributed to the improvement of the paper.

\clearpage

\begin{figure*}
\hspace{2.4cm}
\FigureFile(140mm,60mm){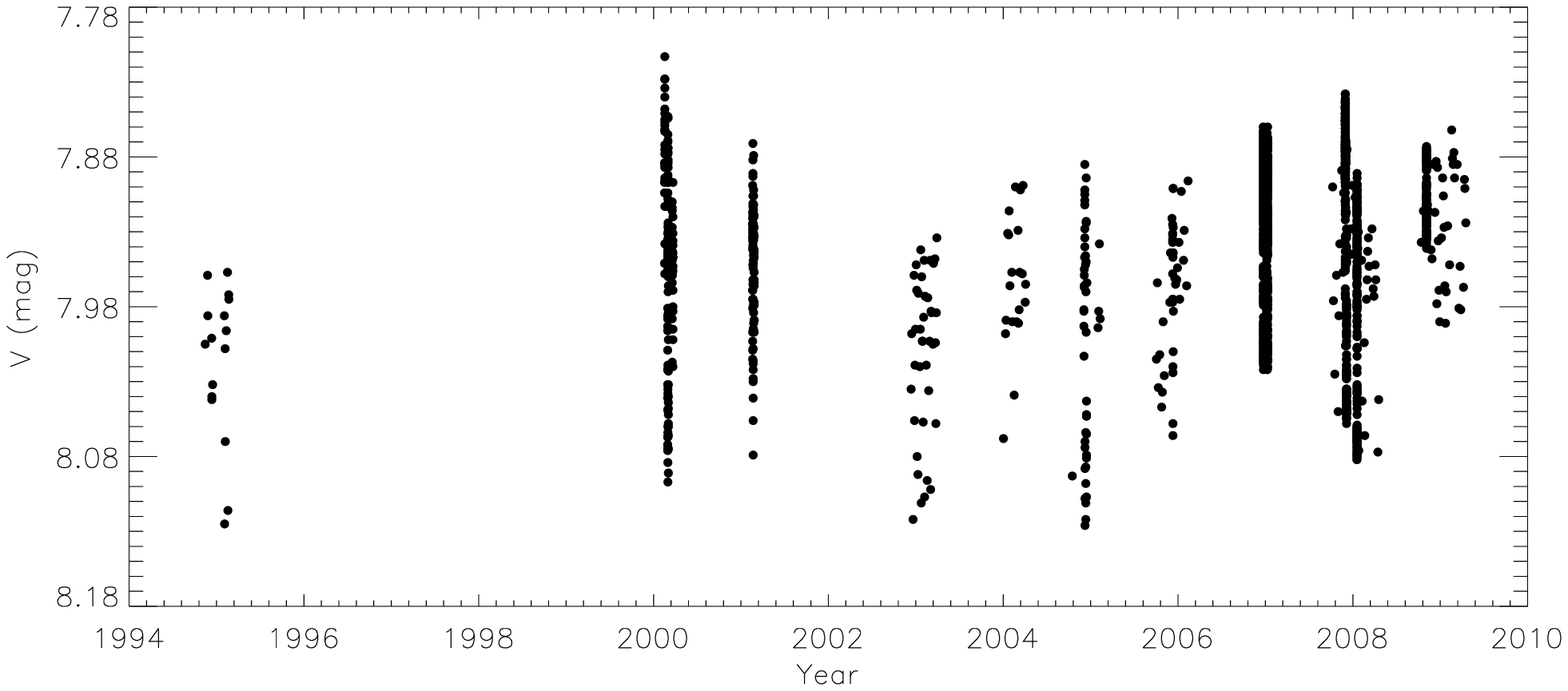}
\vspace{0.5cm}
\caption{All the V-band observations of V369 Gem between November, 1994 and April, 2009.\label{Fig.1}}
\end{figure*}

\begin{figure*}
\hspace{2.0cm}
\FigureFile(160mm,60mm){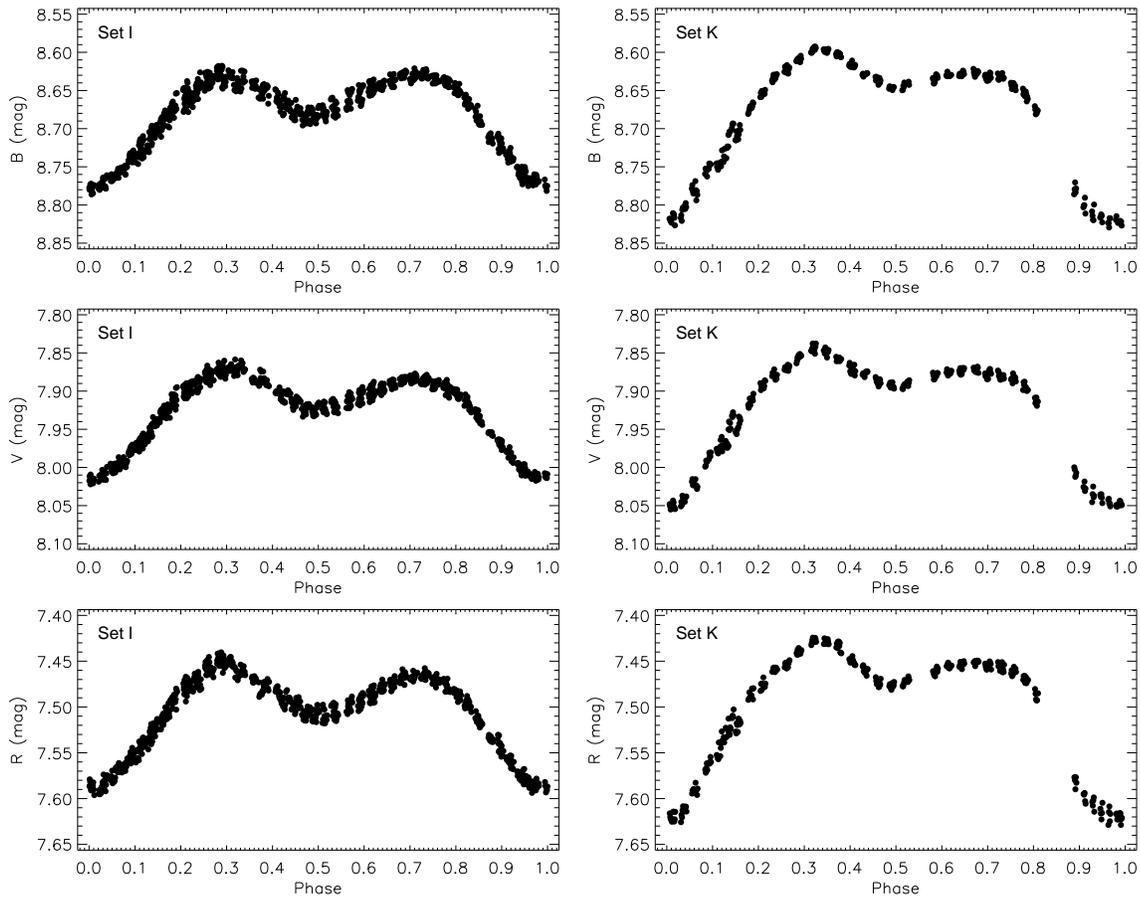}
\vspace{0.5cm}
\caption{The BVR light curves of V369 Gem for different two data sets.\label{Fig.2}}
\end{figure*}

\begin{figure*}
\hspace{5.0cm}
\FigureFile(175mm,60mm){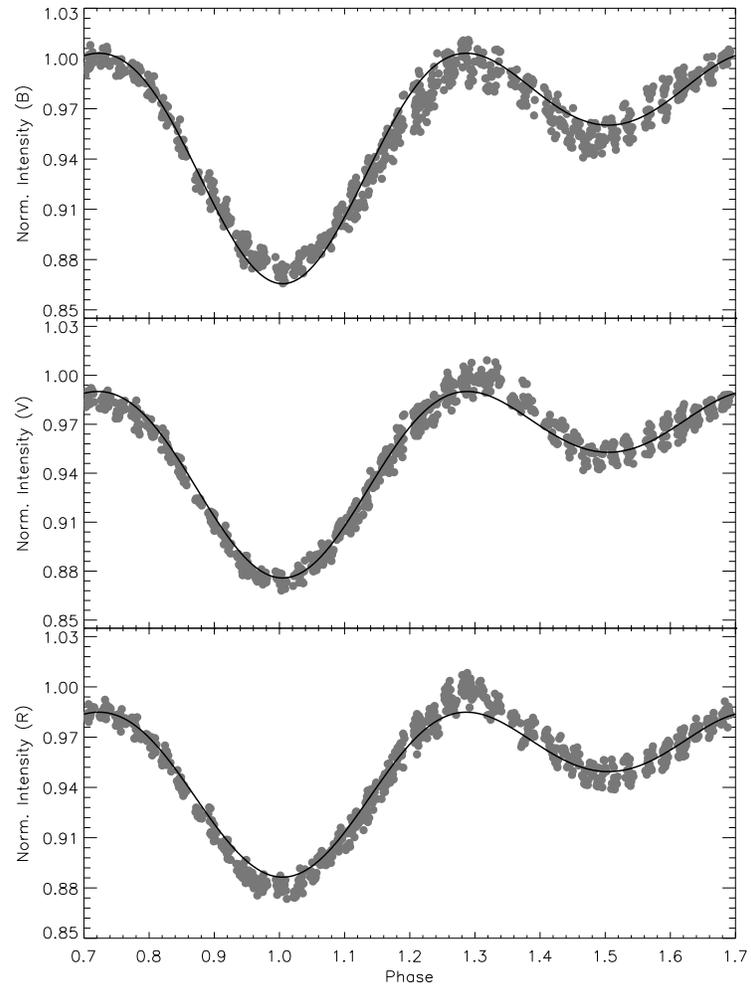}
\vspace{0.5cm}
\caption{The synthetic light curve obtained in just the "overcontact binary not in thermal contact" mode from the light curve analysis.\label{Fig.3}}
\end{figure*}

\begin{figure*}
\hspace{6.5cm}
\FigureFile(55mm,60mm){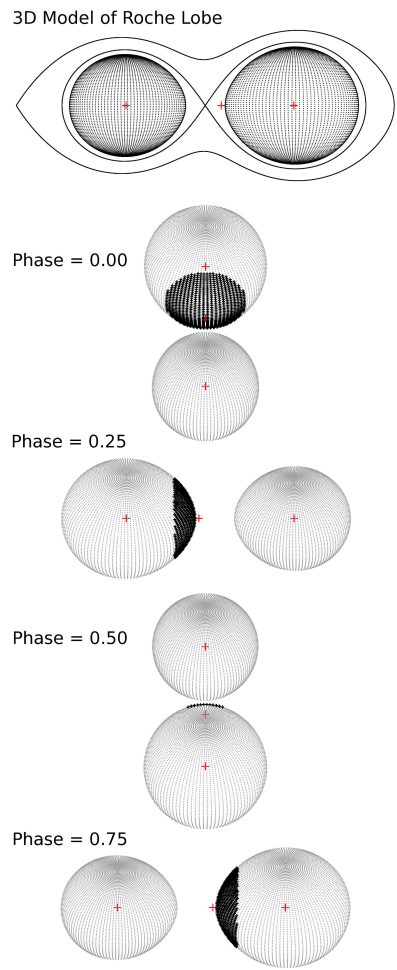}
\vspace{0.5cm}
\caption{The 3D model of Roche geometry and the geometric configurations at four special phases 0.00, 0.25, 0.50 and 0.75, illustrated for V369 Gem by using the PHOEBE V.0.31a software,
using the parameters obtained in just the "overcontact binary not in thermal contact" mode from the light curve analysis.\label{Fig.4}}
\end{figure*}

\begin{figure*}
\hspace{6.0cm}
\FigureFile(155mm,60mm){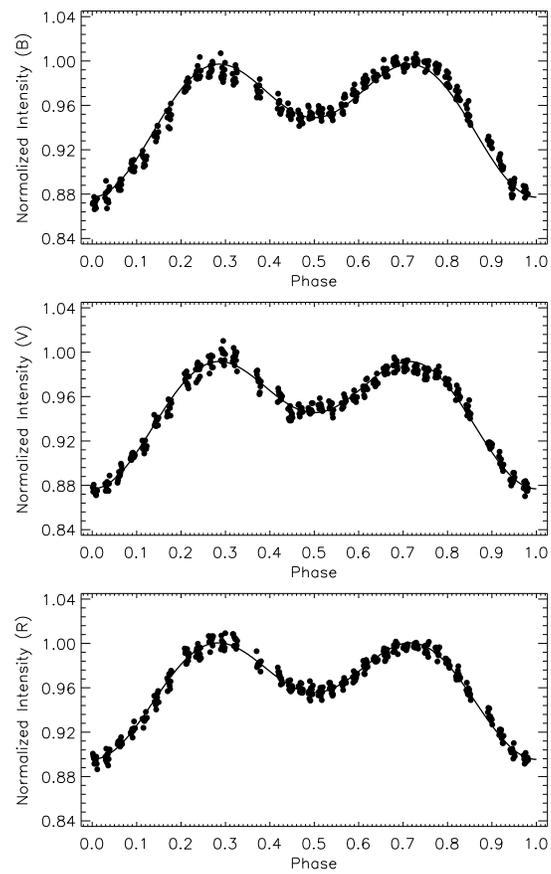}
\vspace{0.5cm}
\caption{BVR light curves of the system obtained in January 10, 2007 (filled circle) and the Fourier models (line).\label{Fig.5}}
\end{figure*}

\begin{figure*}
\hspace{0.90cm}
\FigureFile(185mm,60mm){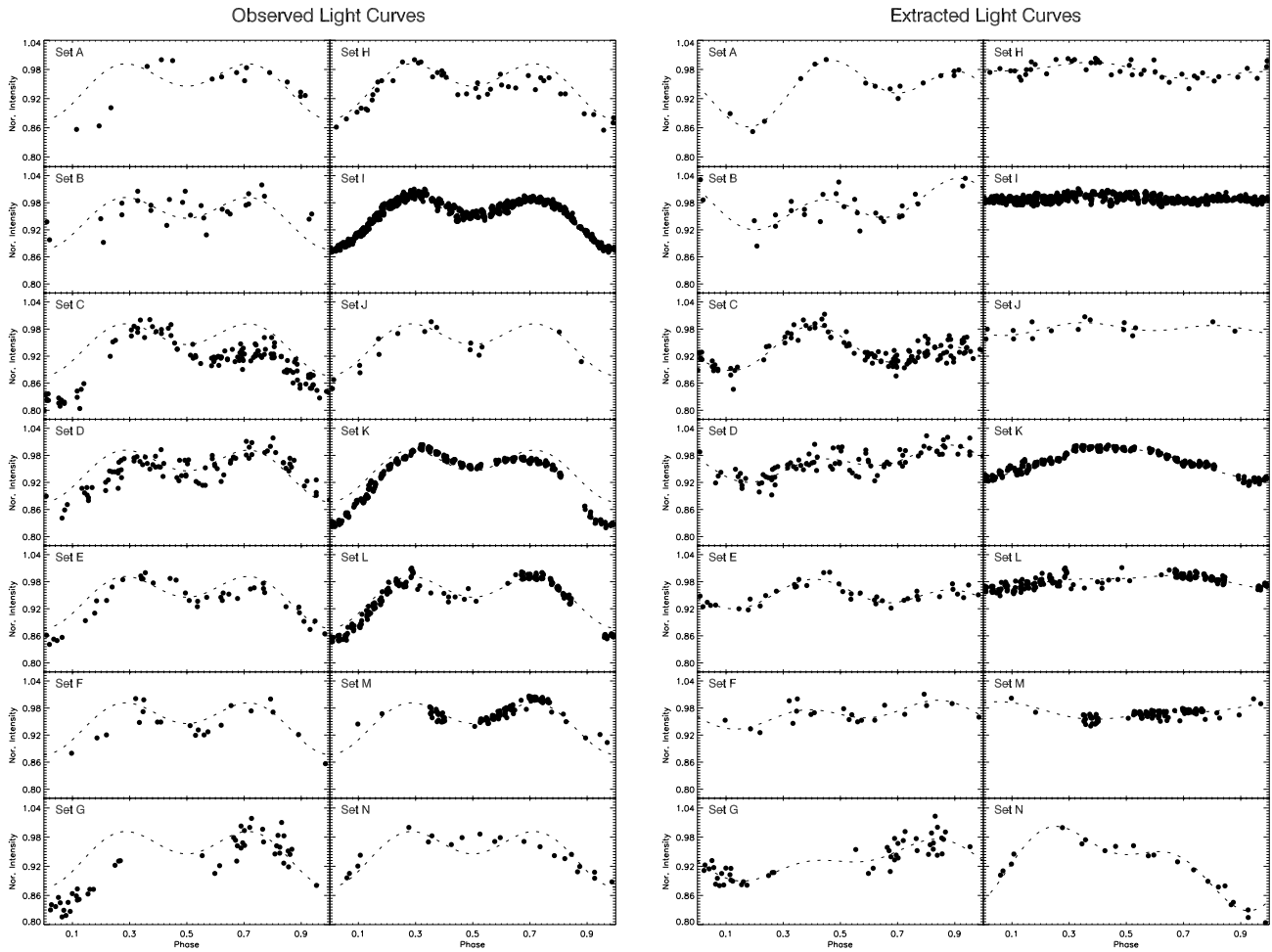}
\vspace{0.5cm}
\caption{V-band light curves for each subset. The curves seen in the first two columns of panels on the left side are observed light curves, while the curves in the last two columns of panels on the right side are the pre-whitened light curves. In the first two columns of the panels, the dashed line in each panel represents the theoretical light curve derived for just the ellipticity effect. However, in the last two columns of the panels, the dashed line in each panel represents the theoretical light curve, which contains just the variation caused by the stellar spots. All light curves are plotted in the same scale to demonstrate the variations of the amplitudes and the shapes.\label{Fig.6}}
\end{figure*}

\begin{figure*}
\hspace{5.7cm}
\FigureFile(140mm,60mm){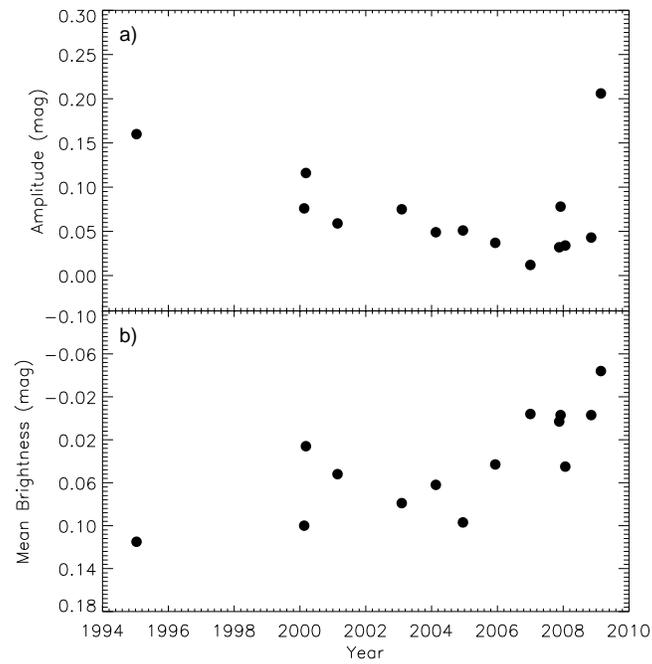}
\vspace{0.5cm}
\caption{The variations of the amplitude (a) and the mean brightness (b) of the pre-whitened light curves for the V-band.\label{Fig.7}}
\end{figure*}

\begin{figure*}
\hspace{2.6cm}
\FigureFile(140mm,60mm){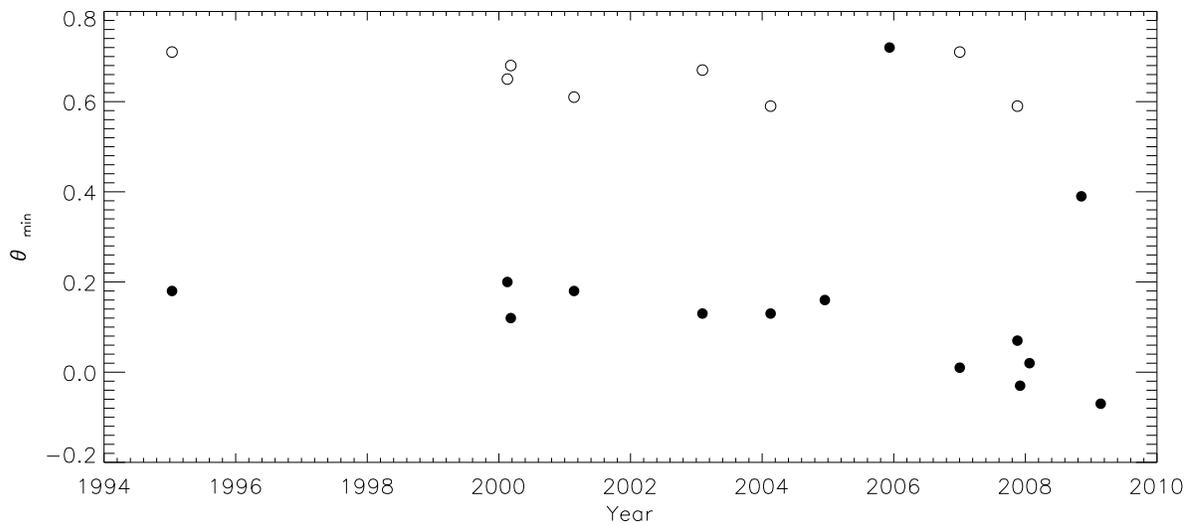}
\vspace{0.5cm}
\caption{The phases of the spot minima ($\theta_{min}$). In figure, filled circle represents Min I, open circles represent Min II.\label{Fig.8}}
\end{figure*}

\clearpage

\begin{table*}
\begin{center}
\caption{Basic parameters of program stars.}
\vspace{0.3cm}
\begin{tabular}{lccc}
\hline\hline						
Stars	&	RA / DE (J2000)	&  V	&	B-V		\\
	&	($^{h}$ $^{m}$ $^{s}$) / ($^\circ$ $^{\prime}$ $^{\prime\prime}$)	& (mag)	&	mag)	\\
\hline							
V369 Gem (HD 52452)   &	07 02 23.32 +25 50 45.56	 &	7.931	&	0.755		\\
HD 52071 (Comparison)	&	07 00 58.15 +27 09 26.50	  &	 7.098	&	1.231	\\
HD 51530 (39 Gem, Check)	&	06 58 47.41 +26 04 51.89	  &	 6.198	&	0.488	\\
\hline						
\end{tabular}
\end{center}
\end{table*}

\begin{table*}
\begin{center}
\caption{Data groups used in the analyses.}
\vspace{0.3cm}
\begin{tabular}{ccccc}
\hline\hline									
Data	&	HJD Interval	&	Median	&	Number	&	Source	\\
Sets	&	(+24 00000)	&	Epoch	&	of Points	&	Ref.	\\
\hline								
 A	&	49672.6213-49772.4665	&	1995.03	&	16	&	1	\\
 B	&	51592.1796-51592.1484	&	2000.13	&	28	&	2	\\
 C	&	51605.1250-51627.2578	&	2000.18	&	129	&	2	\\
 D	&	51963.1992-51963.1796	&	2001.14	&	93	&	2	\\
 E	&	52622.7068-52729.5132	&	2003.09	&	40	&	3	\\
 F	&	53007.7307-53101.4961	&	2004.13	&	21	&	3	\\
 G	&	53295.8523-53412.5385	&	2004.95	&	50	&	3	\\
 H	&	53647.8843-53779.5258	&	2005.94	&	45	&	3	\\
 I	&	54094.2496-54111.6414	&	2007.00	&	889	&	4	\\
 J	&	54383.8712-54465.7459	&	2007.88	&	16	&	3	\\
 K	&	54437.3129-54442.3987	&	2007.92	&	333	&	4	\\
 L	&	54477.7262-54576.4741	&	2008.06	&	209	&	3, 4	\\
 M	&	54753.8753-54830.6945	&	2008.85	&	120	&	3, 4	\\
 N	&	54838.6699-54941.4836	&	2009.15	&	22	&	3	\\
\hline								
\end{tabular}
\end{center}
$^{1}$ \citet{Mes01}. \\
$^{2}$ \citet{Bar04}. \\
$^{3}$ The ASAS database, \citet{Poj97}. \\
$^{4}$ This Study. \\
\end{table*}

\begin{table*}
\begin{center}
\caption{The parameters of the components obtained from the light curve analysis in the "overcontact binary not in thermal contact" mode.}
\begin{tabular}{lr}
\hline\hline
Parameter	&	Value	\\
\hline			
$q$	&	0.76	\\
$i$ ($^\circ$)	&	44.25$\pm$0.12	\\
$T_{1}$ (K)	&	5637 (Fixed)	\\
$T_{2}$ (K)	&	4762$\pm$72	\\
$\Omega_{1}$	&	3.56895	\\
$\Omega_{2}$	&	3.56895	\\
$L_{1}/L_{T}$ (B)	&	0.763$\pm$0.019	\\
$L_{1}/L_{T}$ (V)	&	0.807$\pm$0.021	\\
$L_{1}/L_{T}$ (R)	&	0.728$\pm$0.016	\\
$g_{1}$, $g_{2}$	&	0.32, 0.32	\\
$A_{1}$, $A_{2}$	&	0.50, 0.50	\\
$x_{1,bol}$, $x_{2,bol}$	&	0.646, 0.646	\\
$x_{1,B}$, $x_{2,B}$	&	0.768, 0.768	\\
$x_{1,V}$, $x_{2,V}$	&	0.841, 0.841	\\
$x_{1,R}$, $x_{2,R}$	&	0.677, 0.677	\\
$<r_{1}>$	&	0.3778$\pm$0.0001	\\
$<r_{2}>$	&	0.3285$\pm$0.0001	\\
$Co-Lat_{Spot}$ ($^{\circ}$)	&	90.00	\\
$Co-Long_{Spot}$ ($^{\circ}$)	&	0.00	\\
$R_{Spot}$ ($^{\circ}$)	&	40	\\
$T_{eff,~Spot}$	&	0.93	\\
\hline
\end{tabular}
\end{center}
\end{table*}

\begin{table*}
\begin{center}
\caption{The coefficients derived from the Fourier model.}
\vspace{0.3cm}
\begin{tabular}{ccccccc}
\hline\hline											
Filter	&	$A_{0}$	&	$A_{1}$	&	$A_{2}$	&	$B_{1}$	&	$B_{2}$	\\
\hline										
B	&	0.9531$\pm$0.0003	&	-0.0358$\pm$0.0004	&	-0.0400$\pm$0.0004	&	-0.0062$\pm$0.0004	&	-0.0016$\pm$0.0004	\\
V	&	0.9494$\pm$0.0003	&	-0.0344$\pm$0.0004	&	-0.0383$\pm$0.0004	&	0.0006$\pm$0.0004	&	-0.0026$\pm$0.0004	\\
R	&	0.9615$\pm$0.0002	&	-0.0301$\pm$0.0003	&	-0.0358$\pm$0.0003	&	0.0004$\pm$0.0003	&	-0.0018$\pm$0.0003	\\
\hline										
\end{tabular}
\end{center}
\end{table*}

\begin{table*}
\begin{center}
\caption{The parameters calculated from the pre-whitened light curves.}
\vspace{0.3cm}
\begin{tabular}{cccccccccc }								
\hline\hline											
Data	&	$\theta$	&	$\theta$	&	Amp. (mag)	&	Amp. (mag)	&	Mean V	\\
Set	&       Min I	&    Min II   	&	Min I 	&	Min II	&	(mag)	\\
\hline											
A	&	0.18	&	0.71	&	0.160	&	0.076	&	0.115	\\
B	&	0.20	&	0.65	&	0.076	&	0.044	&	0.100	\\
C	&	0.12	&	0.68	&	0.116	&	0.093	&	0.026	\\
D	&	0.18	&	0.61	&	0.059	&	0.016	&	0.052	\\
E	&	0.13	&	0.67	&	0.075	&	0.059	&	0.079	\\
F	&	0.13	&	0.59	&	0.049	&	0.020	&	0.062	\\
G	&	0.16	&		&	0.051	&		&	0.097	\\
H	&	0.72	&		&	0.037	&		&	0.043	\\
I	&	0.01	&	0.71	&	0.012	&	0.012	&	-0.004	\\
J	&	0.07	&	0.59	&	0.032	&	0.020	&	0.003	\\
K	&	0.97	&		&	0.078	&		&	-0.003	\\
L	&	0.02	&		&	0.034	&		&	0.045	\\
M	&	0.39	&		&	0.043	&		&	-0.003	\\
N	&	0.93	&		&	0.206	&		&	-0.044	\\
\hline															
\end{tabular}
\end{center}
\end{table*}

\begin{table*}
\begin{center}
\caption{The period found from the pre-whitened V light curves.}
\vspace{0.3cm}
\begin{tabular}{cccc}
\hline\hline
Set	&	Year	&	Period (day)			&	FAP ($\%$)	\\
\hline
 A	&	1995.03	&	0.421	$\pm$	0.001	&	4	\\
 B	&	2000.13	&	0.496	$\pm$	0.014	&	57	\\
 C	&	2000.18	&	0.428	$\pm$	0.006	&	2	\\
 D	&	2001.14	&	0.421	$\pm$	0.013	&	8	\\
 E	&	2003.09	&	0.423	$\pm$	0.001	&	8	\\
 F	&	2004.13	&	0.438	$\pm$	0.001	&	8	\\
 G	&	2004.95	&	0.401	$\pm$	0.002	&	6	\\
 H	&	2005.94	&	0.423	$\pm$	0.001	&	2	\\
 I	&	2007.00	&	0.423	$\pm$	0.001	&	2	\\
 J	&	2007.88	&	0.412	$\pm$	0.001	&	5	\\
 K	&	2007.92	&	0.466	$\pm$	0.009	&	2	\\
 L	&	2008.06	&	0.402	$\pm$	0.003	&	2	\\
 M	&	2008.85	&	0.412	$\pm$	0.001	&	2	\\
 N	&	2009.15	&	0.418	$\pm$	0.001	&	6	\\
\hline
\end{tabular}
\end{center}
\end{table*}

\end{document}